# Localized exciton states in π-conjugated polymers with finite torsion

I. Avgin[a], M. Winokur[b] and D. L. Huber[b]

[a] Department of Electrical and Electronic Engineering,

Ege University, Bornova 5100, Izmir, Turkey

[b] Department of Physics, University of Wisconsin-Madison,

Madison, WI 53706, USA

## Abstract

We investigate localized exciton states in π-conjugated polymers with finite torsion. The localized states are associated with a perturbed transfer integral for which the magnitude of cosine of the torsion angle exceeds the magnitude of the corresponding cosine of the unperturbed system. The localized state energy is calculated as a function of the ratio of the perturbed to unperturbed transfer integrals. Particular attention is paid to the optically active symmetric localized states, and the effective oscillator strength, or square of the absolute magnitude of the transition dipole moment, is calculated as a function of the energy. The relation of the theory to recent optical studies of poly(di-n-octylfluorene) (PF8) is discussed.

*Key words: polymer, π-conjugation, torsion, exciton, localized state*

# I. Introduction

Recently, there has been increasing interest in the optical properties of polymers both from a fundamental perspective and for possible application in optical and optoelectronic devices. Among the various classes of polymers, π-conjugated systems are particularly noteworthy for their luminescent properties. In many instances, the low-lying optically active states are excitonic in character involving the $\pi_z$ and $\pi_z^*$ orbitals. If the coupling to the lattice vibrations is relatively weak, one can analyze the excitons in terms of a tight-binding or Frenkel model in which there is a periodic array of identical planar units (platelets) coupled by a transfer integral that is proportional to the cosine of the difference in orientation angles between adjacent platelets [1]. The Frenkel Hamiltonian corresponding to this picture has the form [1,2]

$$H = \sum_j e_j c_j^* c_j - \sum_j t_{j,j+1}(c_{j+1}^* c_j + c_j^* c_{j+1}) \qquad (1)$$

where $e_j$ denotes the optical transition energy of the j[th] platelet, $t_{j,j+1}$ is the transfer integral and the $c_j$ and $c_j^*$ are exciton operators. In our analysis, we will assume that all of the $e_j$ are identical.

As noted, the transfer integrals are proportional to the cosines of the difference in orientation angles of adjacent platelets. In ideal polymers with *torsion* there is a fixed value for the magnitude of the difference. Denoting the torsion angle by $\Delta\varphi_0$ one has $t_{j,j+1} = t\cos(\Delta\varphi_0)$. With a common value for $|\Delta\varphi_0|$ one has a perfectly ordered array. In such a situation, the translational symmetry can be exploited to obtain the exciton energies, which take the form

$$E_k = e_0 - 2t\cos(\Delta\varphi_0)\cos(k) \qquad (2)$$

where $e_0$ is the common transition energy and $k$ ranges between $-\pi$ and $\pi$.

As discussed in Ref. 1, large, random changes in the relative orientation of adjacent planar units are a common source of disorder in the conjugated polymers. In the Frenkel model the changes in angle lead to large changes in the transfer integral. In this paper we will investigate the effect of isolated changes in the transfer integral on the excitonic spectrum. In particular we will show that if there is a pair of adjacent sites where the torsion angle changes from $\Delta\varphi_0$ to $\Delta\varphi_1$ the effect of this change is to create a pair of localized exciton states whenever $|\cos(\Delta\varphi_1)/\cos(\Delta\varphi_0)| > 1$. We will calculate the energies of the localized states, their spatial extent, and their relative oscillator strengths as a function of the ratio $\cos(\Delta\varphi_1)/\cos(\Delta\varphi_0)$. The details of the mathematical analysis will be developed and presented in the following section, while significance of our results and their relation to recent experimental studies of poly(di-n-octylfluorene) (PF8) will be discussed in Sec. III.

## II. Analysis

Our approach to the localized state problem follows is similar to the approach followed in the analysis of localized magnon states [3]. The energy scale in the calculations is in units of the transfer integral in the unperturbed array. The zero of energy is taken to be the center of the exciton band so that in the ideal case the exciton energies range form $-2$ to $+2$. We consider an array of $N$ sites ($N \gg 1$); with $N$ odd, the sites range from $-(N-1)/2$ to $(N-1)/2$. The perturbed transfer integral, $t_{imp}$, connects sites 0 and 1 and has the value $\cos(\Delta\varphi_1)/\cos(\Delta\varphi_0)$.

In the approach of Ref. 3, the eigenstate of the system in the presence of the perturbed transfer integral is expanded in terms of the single-site states $|n>$, corresponding to the excitation of the $n^{th}$ planar unit in the *unperturbed* array. That is, we represent the eigenstate $\Psi$ by a $N$-component column vector

$$|\Psi> = \sum_n |n> \lambda_n \qquad (3)$$

The Schrodinger equation takes the form

$$(H_0 + H_1)|\Psi> = E|\Psi> \qquad (4)$$

where $H_0$ is the Hamiltonian of the unperturbed system, with matrix elements

$$(H_0)_{n,n+1} = (H_0)_{n,n-1} = -1 \qquad (5)$$

and $H_1$ is the perturbation matrix, which has only two non-vanishing matrix elements

$$(H_1)_{01} = (H_1)_{10} = -(t_{imp} - 1) \qquad (6)$$

Rather than work directly with the Schrodinger equation, it is convenient to rewrite it in terms of Green functions for the unperturbed lattice by multiplying it by the operator $G_0 = (E - H_0)^{-1}$. As a result, we obtain the equation

$$(I - G_0 H_1)|\Psi> = 0 \qquad (7)$$

where $I$ denotes the unit matrix. The matrix elements of $G_0$ in the site representation depend only on the magnitude of the separation between the two sites. We have

$$<n|G_0|m> \equiv G_0^{|n-m|} = \pi^{-1} \int_0^\pi d\theta \cos[(n-m)\theta]/[E + 2\cos\theta]$$

$$= (1/E)(1 - 4/E^2)^{-1/2}[(E/2)\{(1 - 4/E^2)^{1/2} - 1\}]^{|n-m|} \qquad (8)$$

for $|E| > 2$, in other words, for energies outside the exciton band of the unperturbed system.

Since the perturbation is limited to the 01 bond, the matrix $G_0 H_1$ has non-vanishing elements only for the columns $n = 0$ and 1. Because of this, the calculation of the localized state energies reduces to a $2 \times 2$ matrix equation of the form

$$(1 - G_0^1(E)H_{10})\alpha - G_0^0(E)H_{10}\beta = 0$$
$$-G_0^0(E)H_{10}\alpha + (1 - G_0^1(E)H_{10})\beta = 0 \qquad (9)$$

The determinant of this set of equations gives the eigenvalues or localized state energies. For each value of $|t_{imp}|>1$, we obtain two localized states, symmetric and antisymmetric with respect to the perturbed bond, having energies $\pm |E_b|$, where $|E_b|$ is given by (see Fig. 1)

$$|E_b| = |t_{imp}| + 1/|t_{imp}| \qquad (10)$$

If $t_{imp} > 1$, the state below the bottom of the exciton band, ($E_b < -2$) is symmetric ($\alpha = \beta$) while the state above the top of the band ($E_b > 2$) is antisymmetric ($\alpha = -\beta$). When $t_{imp} < -1$, the situation is reversed; the state below the bottom of the band is antisymmetric and the state above the top of the band is symmetric.

The (unnormalized) eigenvectors are obtained from the general equation (7). In particular, we have

$$\lambda_n = c(G_0^n(E_b) \pm G_0^{n-1}(E_b)) \qquad n = 1, 2, \ldots \qquad (11)$$

where $c$ is a normalization constant, and the signs denote the symmetric (+) and antisymmetric (−) states.

The spatial extent or localization length of the states depends only on $|E_b|$ and diverges as $|E_b| \to 2$. This can be seen from the form of the Green function (Eq. (8)). The dependence on the factor $[(E/2)\{(1-4/E^2)^{1/2} - 1\}]^{|n-m|}$ translates into exponential decay for the magnitude of the localized state amplitudes. The $|\lambda_n|$ ($n > 0$) decay as $\exp(-n/\delta)$ with $\delta$, identified as the localization length, given by

$$\delta = -[\ln\{(|E_b|/2)(1-(1-4/E_b^2)^{1/2})\}]^{-1} \tag{12}$$

in units of the polymer lattice constant (see Fig. 2). Note that as $|E_b| \to 2$, $\delta$ diverges as $-[\ln(1-(\varepsilon/2)^{1/2})]^{-1}$ with $\varepsilon = |E_b| - 2$.

In the case of optical absorption and emission, only the symmetric states are optically active in lowest order. The relative intensity of the transition is determined by the *effective oscillator strength* or the square of the magnitude of the transition dipole matrix element. For normalized wave functions ($\sum_n |\lambda_n|^2 = 1$), an ideal array of $N$ chromaphores with periodic boundary conditions has all the intensity in the $k = 0$ mode for which the effective oscillator strength is $N$. In general, the effective oscillator strength (*EOS*) for a localized mode is defined by the sum

$$EOS = |\sum_n \lambda_n|^2 \tag{13}$$

where the $\lambda_n$ are components of the normalized eigenvector for the mode.

We have calculated the effective oscillator strength for the symmetric modes for the case $t_{imp} > 1$ (symmetric state below the bottom of the band). Denoting the effective oscillator strength in this case by the symbol $R$ we found

$$R = 2[(|E_b|+2)/(|E_b|-2)]^{1/2} \tag{14}$$

When $t_{imp} < -1$ (symmetric state above the top of the band), we obtained a different result for the effective oscillator strength (denoted by $S$)

$$S = 2[(|E_b|-2)/(|E_b|+2)]^{1/2} \tag{15}$$

In Figs. 3 and 4, the effective oscillator strengths are plotted as a function of the energy $|E_b|$. It is evident that there is a qualitative difference between the two figures.

This difference in behavior reflects the fact that the wave function characterizing the symmetric localized state lying below the bottom of the band ($t_{imp} > 1$) decreases monotonically with increasing distance from sites 0 and 1. In contrast, the wave function for the symmetric localized state above the top of the band ($t_{imp} < -1$) oscillates in sign with increasing distance from the perturbed bond. It is the oscillation that reduces the magnitude of the *EOS*. It should be noted that there are no localized states when $|t_{imp}| \leq 1$. In this case the impurity introduces an extended "resonant mode" that is degenerate with the exciton band [3].

### III. Discussion

In the preceding section, we have presented the results of a calculation of the localized exciton states in a π-conjugated polymer with torsion. The localized states arise from a transfer integral characterized by an abrupt change in torsion angle. If the torsion angle of the unperturbed chain is $\Delta\varphi_0$ and the perturbed angle is $\Delta\varphi_1$, the condition for a localized state is $|\cos(\Delta\varphi_1)/\cos(\Delta\varphi_0)| > 1$. For a normal band ($k = 0$ state is lowest in energy), when $\cos(\Delta\varphi_1)/\cos(\Delta\varphi_0) > 1$, the symmetric (optically active) localized state lies below the bottom of the band at the energy $-|E_b|$, and the antisymmetric localized state lies above the top of the band at the energy $|E_b|$. When $\cos(\Delta\varphi_1)/\cos(\Delta\varphi_0) < -1$, the symmetric state lies above the top of the band at the energy $|E_b|$ while the antisymmetric localized state lies below the bottom of the band at energy $-|E_b|$. In the case of an inverted band ($k = 0$ state is highest in energy), the situation is reversed. When $\cos(\Delta\varphi_1)/\cos(\Delta\varphi_0) > 1$, the symmetric b state lies above the top of the band at the energy $|E_b|$ and the antisymmetric localized state lies below the bottom of the band at energy $-|E_b|$. When $\cos(\Delta\varphi_1)/\cos(\Delta\varphi_0) < -1$, the symmetric state lies below the band at the energy $-|E_b|$ while the antisymmetric localized state lies above the top of the band at energy $|E_b|$.

We have also determined the *EOS* for both classes of symmetric localized state. It was noted previously that symmetric localized states with energies near the energy of the $k = 0$ mode of the unperturbed array have a larger *EOS* (for the same $|E_b|$) than the localized states with energies near the energy of the $k = \pi$ mode, although both *EOS* approach 2 in the limit $|\cos(\Delta\varphi_1)/\cos(\Delta\varphi_0)| \to \infty$. The difference in the behavior of the *EOS* reflects the fact that the localized state wave functions can be viewed as wave packets of the extended states from the adjacent region of the exciton band. When the adjacent region is near the center of the Brillouin zone, the localized state amplitudes, $\lambda_n$, vary slowly (similar to the modes with $k \approx 0$) and hence have a large *EOS*, whereas when the localized states are near the zone boundary, the amplitudes have an oscillatory character similar to the extended states near the zone boundary which have zero oscillator strength. The behavior as $|\cos(\Delta\varphi_1)/\cos(\Delta\varphi_0)| \to \infty$ is indicative of the fact that in this limit, the amplitudes are confined to the sites on either side of the perturbed bond so that the *EOS* is equal to $(1/2^{1/2} + 1/2^{1/2})^2 = 2$. We have also analyzed the spatial extent of the localized state wave functions. The power law behavior of the Green function leads to exponential decay of the wave function with a decay length that depends only on the localized state energy and diverges as the localized state energy approaches the exciton band of the ideal lattice.

We are aware of at least one system that meets the criteria for optically active localized states. In previous work on poly(di-n-octylfluorene) (PF8) [4,5], it was established that the low-temperature equilibrium state consisted of a mixture of α (90.5%) and β (9.5%) phases characterized by different torsion angles. In the samples studied in Ref. 5, it was found that the torsion angles of the α phase were equal to 136.5°, while the torsion angles of the β phase were equal to 165°.[*] If we view the torsion angles of the β phase as "isolated impurity bonds", the corresponding value $t_{imp}$ is $\cos(165°)/\cos(136.5°) = 1.33$ for which the symmetric localized state energy is −2.08 and the localization length $\delta = 3.55$, both in dimensionless units. Since the transfer integral in the α phase is equal to 0.98 eV [5], the localized state lies below the bottom of the α exciton band by about 0.08 eV. The effective oscillator strength associated with this localized state is equal to 14.3 so that even a small concentration of "β bonds" should have a noticeable effect on the absorption spectrum. The 18 K absorption spectrum of PF8 displayed in Fig. 3 of Ref. 5 shows a peak approximately 0.2 eV below the peak in the 110 K α-spectrum which is qualitatively consistent with an isolated localized state, but further studies are necessary to confirm this hypothesis.

---

[*] Note that the sign of the transfer integral in Ref. 5 is opposite to what is used in this paper (+$t$ instead of −$t$). As a consequence, $\Delta\varphi > \pi/2$ in Ref. 5 is equivalent to a torsion angle $\pi - \Delta\varphi$ in this paper. Thus the α exciton band has the normal form (minimum energy at $k = 0$).

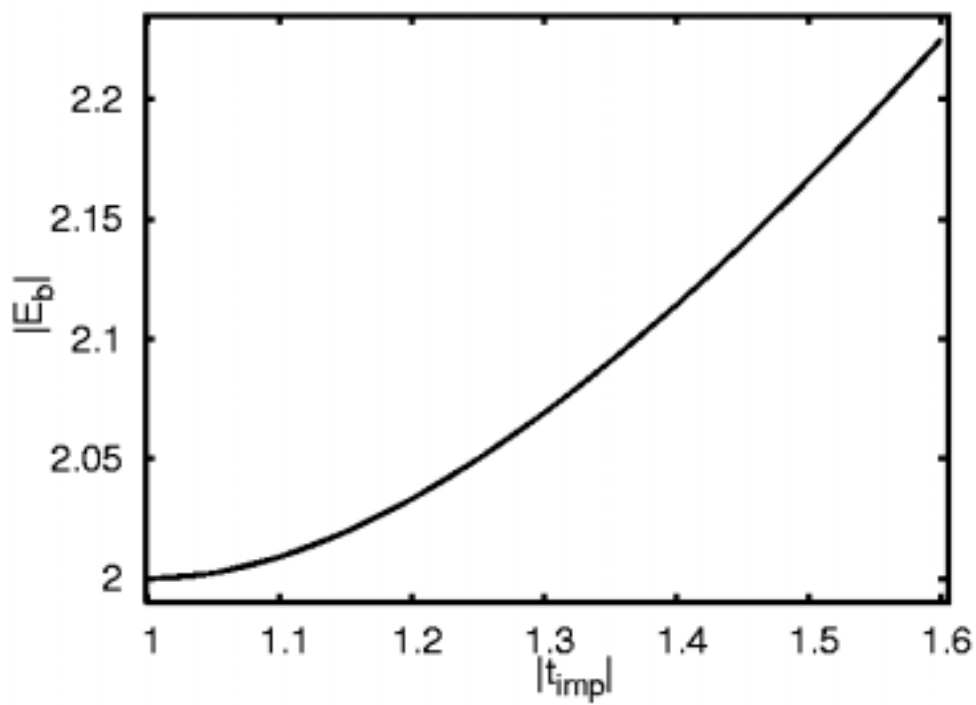

Fig. 1. Magnitude of the localized state energy, $|E_b|$, vs magnitude of $|t_{imp}|$. For $|t_{imp}| > 1$, there are localized states at energies $\pm |E_b|$. The symbol $t_{imp}$ denotes the ratio of the cosine of the torsion angle of the perturbed bond to the cosine of the torsion angle of the unperturbed bond, i.e. $t_{imp} = \cos(\Delta\varphi_1)/\cos(\Delta\varphi_0)$.

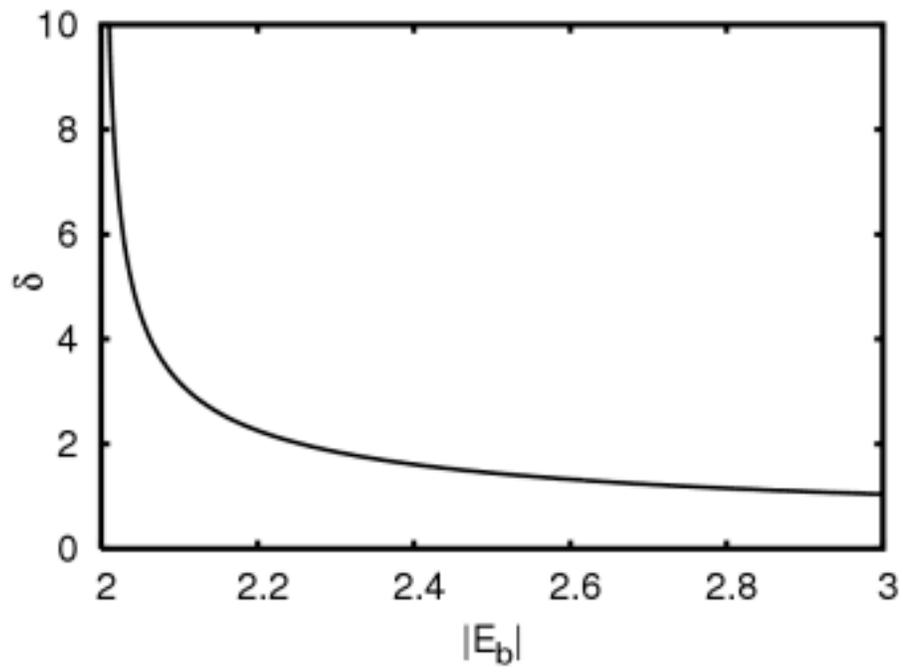

Fig. 2 Exciton localization length, δ, vs |E$_b$|

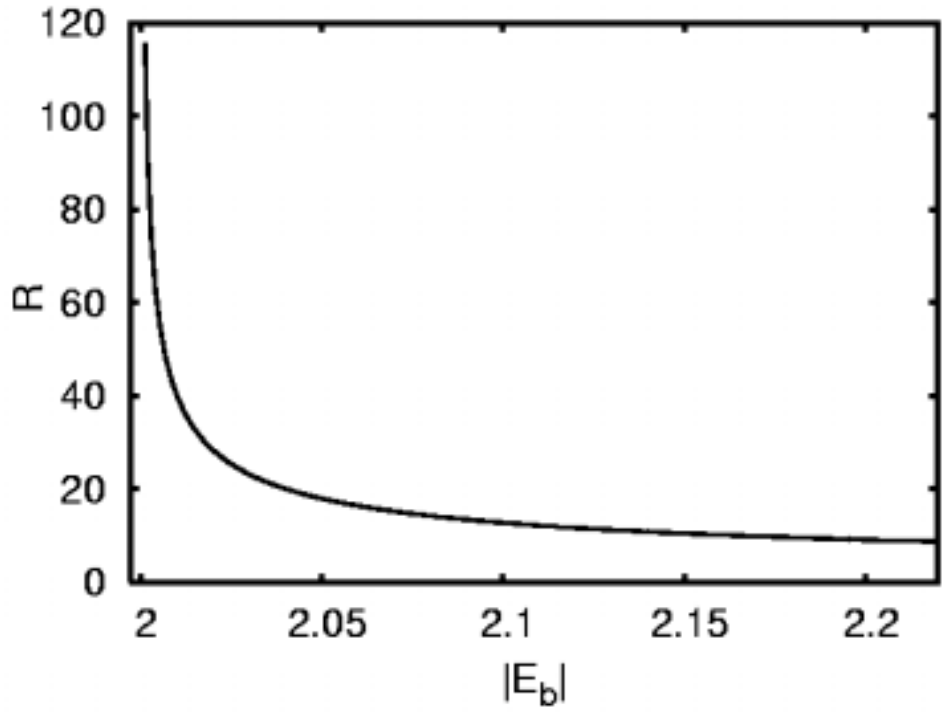

Fig. 3. Effective oscillator strength for $t_{imp} > 1$, denoted by $R$, vs $|E_b|$. In this case, the optically active, symmetric mode lies below the bottom of the unperturbed exciton band ($E_{k=0} = -2$).

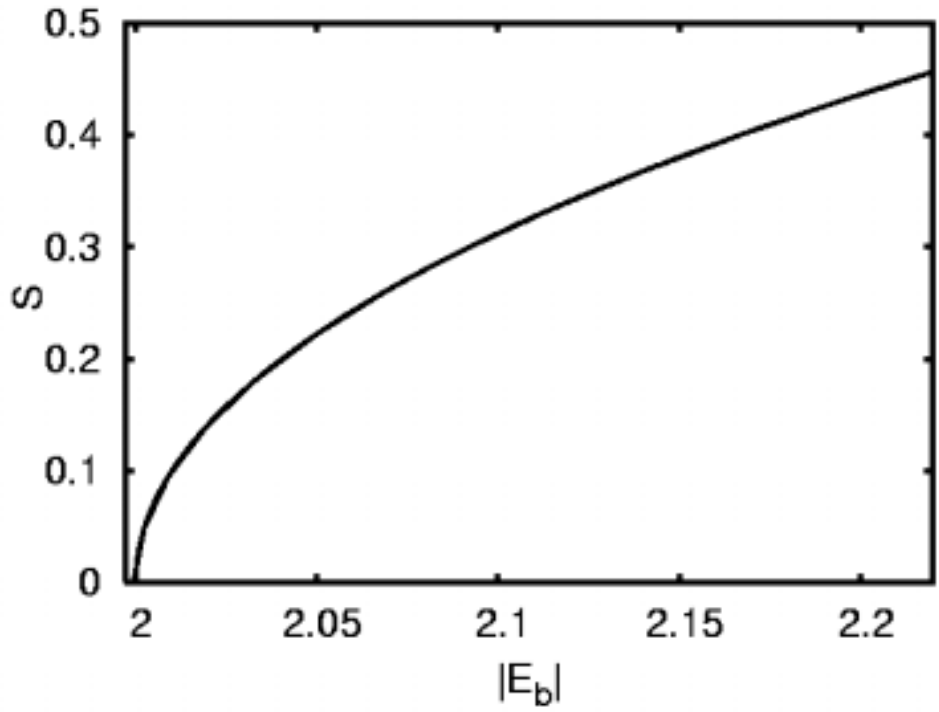

Fig. 4. Effective oscillator strength for $t_{imp} < -1$, denoted by $S$, vs $|E_b|$. In this case, the optically active, symmetric mode lies above the top of the unperturbed exciton band ($E_{k=0} = -2$).